\documentstyle[twocolumn]{article}

\begin{document}

\twocolumn[%
\title{
Topological entropy of a stiff ring polymer and its connection to  DNA knots }

\author{
Miyuki K. {\sc Shimamura}
 and Tetsuo {\sc Deguchi} \\
\\
Department of Physics, Faculty of Science \\ 
and Graduate School of Humanities and Sciences, \\
Ochanomizu University \\
2-1-1 Ohtsuka, Bunkyo-ku, Tokyo 112-8610, Japan\\
\\
E-mail: miyuki@degway.phys.ocha.ac.jp} 

\date{}

\maketitle

\begin{abstract}
We discuss  the entropy of a circular polymer 
under a topological constraint.  We call it the {\it topological entropy} 
of the polymer, in short.  
A ring polymer  does not change its topology  (knot type)  
 under any thermal fluctuations.  
 Through numerical simulations using some knot invariants, 
 we show that   
the topological entropy of  a  stiff ring polymer 
with a fixed knot is described by a scaling formula 
as a function of the thickness and length of the circular chain. 
The result is consistent with the viewpoint that 
for stiff polymers such as DNAs,  
the length and diameter of the chains should play 
a central role in their statistical 
and dynamical properties.  Furthermore, 
we  show that the new formula extends 
a known theoretical formula for DNA knots.  
\end{abstract}

\vspace{3mm}
$\bf{keyword}$ \quad
topological entropy, circular polymers, 
random knotting, self-avoiding polygons, knot invariants, DNA knots\\

\vspace{1cm}]
 
\setcounter{equation}{0} 
\renewcommand{\theequation}{1.\arabic{equation}}
\setcounter{section}{0} 

\section{Introduction}
In the last fifteen years, 
various knotted ring-polymers are synthesized 
and observed in  experiments of chemistry and biology. 
\cite{Dean,Shishido,Walba}
Once a ring polymer is formed, its topological state, 
which is given by a knot,   is unique and invariant. 
One of unsolved problems in statistical physics of macromolecules is how 
to formulate the topological constraint on ring polymers theoretically 
so that we can investigate topological effects on their statistical and dynamical properties.
This problem was  addressed by Delbr{\"u}ck in the 60's \cite{Delbruck,Frisch-Wasserman}.
Since then, several numerical simulations have been performed \cite{Vologodskii1,desCloizeaux,Michels,LeBret,Chen,Klenin,Janse,Koniaris,DeguchiJKTR,Deguchi95,DeguchiRevE,Orlandini}.
The topological constraint on a ring-polymer is nontrivial. 
It may severely restrict the available degrees 
of freedom in the configuration space, and then 
 lead to a large reduction on its entropy.

Let us  formulate the topological problem, explicitly. 
We consider a ring polymer in good solution, whose degree of polymerization 
corresponds to  $N$ units of the Kuhn statistical length \cite{Grosberg-book}. 
Then,  a spatial configuration of the ring polymer can be approximated 
by a self-avoiding polygon with $N$ polygonal nodes 
(or $N$ bonds, $N$ vertices).  
Here, the length of polygonal segments is given by the Kuhn length.

Let us assume   the number  $W_K(N)$ 
of all possible $N$-noded self-avoiding polygons with a knot $K$.    
Then,  the topological entropy $S_K(N)$ for the knot $K$ 
is given by  $S_K(N)=k_{\rm B} \log W_K(N)$.  
Here $k_{\rm B}$ is the Boltzmann constant.
We remark that  the entropy $S(N)$ 
 without any topological constraint 
is given by $S(N)=k_{\rm B}\log W(N)$, where 
 $W(N)$ is the number of all self-avoiding polygons
with $N$ nodes: $W(N)=\sum_K W_K(N)$. 
Let us introduce the knotting  probability $P_K(N)$ of a  knot $K$. 
We define it by the probability that 
a given configuration of a self-avoiding polygon with $N$ nodes 
is equivalent to the knot $K$.  It is clear that 
the knotting probability is given  by  $P_K(N)=W_K(N)/W(N)$.
Thus, the decrease  of the entropy of the ring polymer  due to the topological constraint 
that it should keep its knot type $K$ is expressed  in terms of 
the knotting probability for the knot $K$:
\begin{eqnarray}
S(N)-S_K(N) &=& k_{\rm B} \log W(N) -k_{\rm B} \log W_K(N) \nonumber \\
              &=& k_{\rm B} \log(1/P_K(N)).
\end{eqnarray}

\par Recently, randomly knotted DNA rings  are synthesized in experiments,  
and the fractions of 
some knotted species are measured through the electrophoretic separation. \cite{Rybenkov,Shaw}
Here we remark that  the fraction of DNAs with  a knot $K$  corresponds to 
the knotting probability of the  knot $K$. 
The measurement of the fractions of  knotted DNAs 
is  particularly interesting 
 because it gives an excellent method for determining 
the effective diameter of DNAs,  which are negatively charged
and  surrounded by  the clouds of counter ions.  
The effective diameter of DNAs in an electrolyte solution depends 
strongly on the concentration of the counter ions. 
However, it is nontrivial to calculate the effective diameter of DNAs  
by taking into account the electric double layers around the cylindrical segments.  
\cite{Stiger,Brian} 
Furthermore, 
DNA chains are rather stiff;    
the persistent length (i.e., the Kuhn statistical length) 
of DNA chains is   given by 50 $\sim$ 100 nm, 
depending on the ionic concentrations \cite{OSF},  
which is much larger than that of  polystylenes. \cite{Grosberg-book}

\par 
Thermodynamical properties of 
DNA chains in electrolyte solutions 
are studied by introducing 
 some models of cylindrical self-avoiding walks. 
For DNA rings,  we introduce 
a model of self-avoiding polygons  consisting of 
$N$ cylindrical segments with cylindrical radius $r$ where 
their length is  given by the Kuhn  length $b$. 
Then,  let us consider the probability  
$P_{knotted}(N,r)$  of a cylindrical polygon being knotted 
(i.e., being a non-trivial knot).  
 For the algorithms of cylindrical self-avoiding polygons \cite{LeBret,Klenin},    
it was studied through simulations  
how the probability $P_{knotted}(N,r)$ 
should depend on the radius $r$. The numerical results were analyzed   
by assuming an exponential dependence in the following:  
\begin{equation} 
P_{knotted}(N, r) = P_{knotted}(N,0) \exp\left( - \gamma \frac{r}{b} \right) \, .
\label{rb}
\end{equation}
Here $\gamma$ is a constant to be determined by the simulations. 
Thus, by measuring  the fraction of knotted DNAs 
and applying the theoretical formula (\ref{rb}) to them,  
 the effective diameters of DNA rings in some electrolyte solutions 
are determined. \cite{Rybenkov,Shaw}

\par 
In this paper, we introduce a theoretical formula 
describing  the knotting probability of a given knot 
for  stiff ring polymers 
such as circular DNAs. 
It describes how the knotting probability  depends on the length and 
radius of the ring polymers. 
We formulate an algorithm of self-avoiding polygons consisting 
of cylindrical segments with radius $r$ and unit length. \cite{PLA} 
The algorithm is based on the dimerization algorithm, and we call
the algorithm the cylindrical ring-dimerization method, 
or the dimerization method, for short.
 The method should be useful for studying stiff ring polymers. 
In fact, it is closely related to the wormlike chain model  
for polyelectrolytes in electrolyte solutions.  \cite{OSF} 
Through numerical simulations with the method, 
we evaluate the knotting probabilities for the cylindrical 
self-avoiding polygons with several different numbers $N$ 
of polygonal nodes and 
 different values of the cylindrical radius $r$. 
Then, we  show that the new formula gives good fitting 
 curves to the estimates of the knotting probabilities.  
 In fact, the fitting curves fit to the data for the cylindrical polygons   
of very small as well as very large (even asymptotically large) numbers $N$ of nodes, 
and  of all the different values of the radius $r$.  
Furthermore, we show that it extends the known formula (\ref{rb}) 
for the DNA knots. Finally,  
 we discuss some consequences of the theoretical formula of the knotting probability,  
which may be useful for future study  of knotted circular DNAs.

\par 
The model of cylindrical self-avoiding polygons 
should be valid for any real ring polymer in good solution. 
According to the standard two-parameter theory of polymers, 
it is expected that any statistical property of polymers can be described  
by the length and thickness of the chains. \cite{Grosberg-book}   
The radius of cylindrical segments corresponds to the thickness 
of ring polymers, 
which expresses the excluded-volume effect.    
In appendix A, some algorithms of the cylindrical self-avoiding polygons 
are shown.

\par 
The paper is organized as follows:
In \S 2 we explain the cylindrical ring-dimerization method of our numerical experiment of random knotting.
In \S 3, employing the algorithm discussed in \S 2, 
we produce a large number of self-avoiding polygons consisting of cylinders,  
and present numerical estimates of the knotting probabilities  
of several different knots such as the trivial knot, 
non-trivial prime knots ($3_1$, $4_1$, $5_1$, $5_2$), 
and nontrivial composite knots  ($3_1 \sharp 3_1$, $3_1 \sharp 3_1\sharp 3_1$).
Then,  we find that they are well described by the new  
theoretical  formula of the knotting probability of the 
cylindrical self-avoiding polygons constructed by the cylindrical ring-dimerization method.
In \S 4, we show that the new  formula generalizes  the 
known formula (\ref{rb}).   
Finally, we  give some concluding remarks in \S 5.

\setcounter{equation}{0} 
\renewcommand{\theequation}{2.\arabic{equation}}
\setcounter{section}{1} 

\section{Methods of Numerical Simulations}
\subsection{Overview of the numerical experiment of random knotting}

We first discuss the outline of  our numerical experiment.
The knotting probability 
is evaluated by the following processes:  
(1) We construct a large number, say $M$,  of 
 self-avoiding polygons 
of $N$ cylindrical segments with radius $r$ and unit length; 
(2) 
Projecting  the three-dimensional configurations of the self-avoiding polygons 
onto a plane, we make their knot diagrams;  
(3) Calculating some knot invariants   for the knot diagrams, 
we  enumerate the number $M_K$ of such polygons that 
have the same set of values of the knot invariants with a knot $K$. 

For a  given number $N$ of polygonal nodes,  we produce $M$ polygons with the 
algorithm  of the dimerization method of self-avoiding polygons.
In this paper, we set $M=10^4$. 
 We recall that $M_K$ denotes 
 the number of  such polygons 
that have the same set of values of the knot invariants with the knot $K$.
Then, it is clear that the knotting probability $P_K(N)$ is given by the following
\begin{equation}
P_K(N)=\frac{M_K}{M}
\end{equation}
We note that the algorithm of the dimerization method
will be  explained  in \S 2.3 and \S 2.4 (See also \S A.1). 

\subsection{Knot invariants for detecting  knot types}

In the numerical simulations of the paper, 
the tool for detecting the knot type 
of a given polygon is a set of two knot invariants in the following:
the determininant $\Delta_K(-1)$ of knot and 
the Vassiliev-type invariant $\it{v_2}(K)$ of the second degree.
The values of the invariants for some typical knots are given in 
Table \ref{inv}.
It is remarked that the two knot invariants have definite advantages 
for practical purposes. In fact, there exist some algorithms 
by which we can calculate the two knot invariants 
in  polynomial time with respect to the number of crossing points 
of a  given knot diagram.  \cite{DeguchiPLA,Polyak}

\subsection{Cylindrical ring-dimerization method}

Let us discuss the method \cite{PLA}  for constructing  
self-avoiding polygons consisting of cylinders with radius $r$ of unit length.  
It is based on the dimerization algorithm. \cite{Madras} 
We recall that it is called the cylindrical ring-dimerization method,
or the dimerization method, for short. 
The method consists of the following processes:
(1)
 We generate a set of chains with cylindrical segments by the dimerization algorithm;  
(2)  We construct polygons by connecting two cylindrical self-avoiding chains 
with the method of Ref. \cite{Chen}, where we  
also calculate the statistical weight related to the probability of successful concatenation.

Let us explain  the dimerization algorithms for self-avoiding walks,  
and then for self-avoiding polygons. 
Under the dimerization algorithm \cite{Madras},  
 we make a chain by connecting 
two given subchains if they have no ``overlap''; if they have an ``overlap'', 
then we throw away both of the two subchains 
and we take a new pair of subchains to try. 
In the method  \cite{Chen} for  making self-avoiding polygons, 
first we pick up a compatible pair of self-avoiding walks, and then we check 
whether they have any ``overlap'' or not. If not, we make a 
new ring by connecting the ends of the two chains; if there is an ``overlap'', 
then we throw them away and take a new pair of chains. 
In the concatenating method, 
 we also calculate the statistical weight related to the probability of 
  successful concatenation.

One of the key points of our algorithm is the condition  
when  given two cylindrical segments have an ``overlap''. 
We define the condition of `` overlap'' between a given pair of segments 
as follows. 
First, there is no ``overlap'' between any pair of adjacent segments, i.e., 
we neglect the thickness of segments 
for  any pair of adjacent cylindrical segments. Thus, 
there is no constraint between any next-neighboring segments 
in our model of cylindrical self-avoiding polygons. 
Second, two segments have no ``overlap'' 
if the minimum distance between the central line segments 
of cylinders for any given pair of unadjacent cylinders 
is larger than the diameter $2r$ of cylinders.
Here, we have defined the central line segment of a cylinder by 
the line segment between the centers of the upper and lower disks 
of the cylinder.
For an illustration,  we give  in Appendix B 
the algorithm for checking whether any given two cylindrical segments 
have an ``overlap'' or not.

The cylinder condition is important in practical applications. 
We recall that DNA chain as a polyelectrolyte in an electrolyte solution 
can be  modeled by the model of cylindrical self-avoiding polygons since  
the negatively charged DNA segments 
with counter ions shielded around them can be approximated 
effectively  as impermeable cylinders of the Kuhn statistical length with  
the radius given by the Debye screening length.

\subsection{Some details of the dimerization algorithm}

Let us describe the algorithm of dimerization, more precisely. \cite{Madras} 
First, we construct a very large number, say $M_0$, of 
 short self-avoiding chains with $N_0$ cylindrical segments, randomly 
by a direct method.  
Then, we pick up a pair of short chains out of the $M_0$ 
generated chains randomly,  
and see whether there is any overlap or not between the chains; we investigate 
 all unadjacent pairs of cylindrical segments of the chains.
If there is an overlap, then we give up the chains and consider  a new pair 
of short chains from the beginning.
If there is no overlap, then we make a longer chain with length $2 N_0$ 
by connecting the end of one chain to the head of another one. 
After constructing $M_0/2$ chains with length $2 N_0$ by the method, 
we repeat this construction for the $M_0/2$ longer chains. 
The scheme of the dimerization algorithm is shown in Fig. 3, which is given by 
a binary tree.  
Thus, we can construct very long 
self-avoiding polygons with the cylinder radius $r$, quite efficiently.
In appendix A, we explain all the detail of the algorithm of the
cylindrical ring-dimerization, explicitly.

\setcounter{equation}{0} 
\renewcommand{\theequation}{3.\arabic{equation}}
\setcounter{section}{2} 
\section{Knotting Probability and Its Fitting Formula}

\subsection{A brief review on knotting probabilities of ring polymers}

We discuss some important known results on the statistics 
of random knots obtained 
by numerical simulations with algebraic invariants of knots.
In particular, we consider how the knotting probability should depend 
on the length $N$ of ring polymers. 

Let us first consider the knotting probability of the trivial knot. 
We recall that it is the probability of forming a  trivial knot 
in a given random polygon (or self-avoiding polygon) with $N$ nodes.
We  denote it by $P_0(N,r)$, and call it  the unknotting probability or 
 knotting probability of the  trivial knot (or unknot). 
Here, the symbol $K=0$ denotes that the knot $K$ is trivial.
We also note that the probability $P_{knotted}(N,r)$ 
of a polygon being knotted 
is given by  the unknotting probability $P_0(N,r)$ 
as in the following: $P_{knotted}(N,r)=1-P_0(N,r)$. 
Here we recall that if a polygon is equivalent 
to a nontrivial knot, then we regard  it as knotted.

The unknotting probability  has been evaluated 
for different models and methods:  
some models of the closed random polygons 
\cite{Vologodskii1,desCloizeaux,Michels}, the hedge-hog method \cite{Klenin}, 
and the rod-bead model \cite{Chen,Koniaris}. For 
some models of random polygons or self-avoiding polygons, 
it has been found in Ref. \cite{Michels,Koniaris} 
that the unknotting probability  
has a decreasing exponential dependence 
on the number  $N$ of nodes: 
\begin{equation}
\label{tri}
P_0(N)=C_0\exp(-N/N_0) .
\end{equation}
Here, the two parameters $N_0$ and $C_0$ are to be determined so that
eq. (\ref{tri}) gives the best fitting curve  
to the numerical estimates of the unknotting probabilities 
for different numbers $N$. 
We call the parameter $N_0$ the characteristic length 
of random knotting  of unknot.

Let us consider the case of nontrivial knots. 
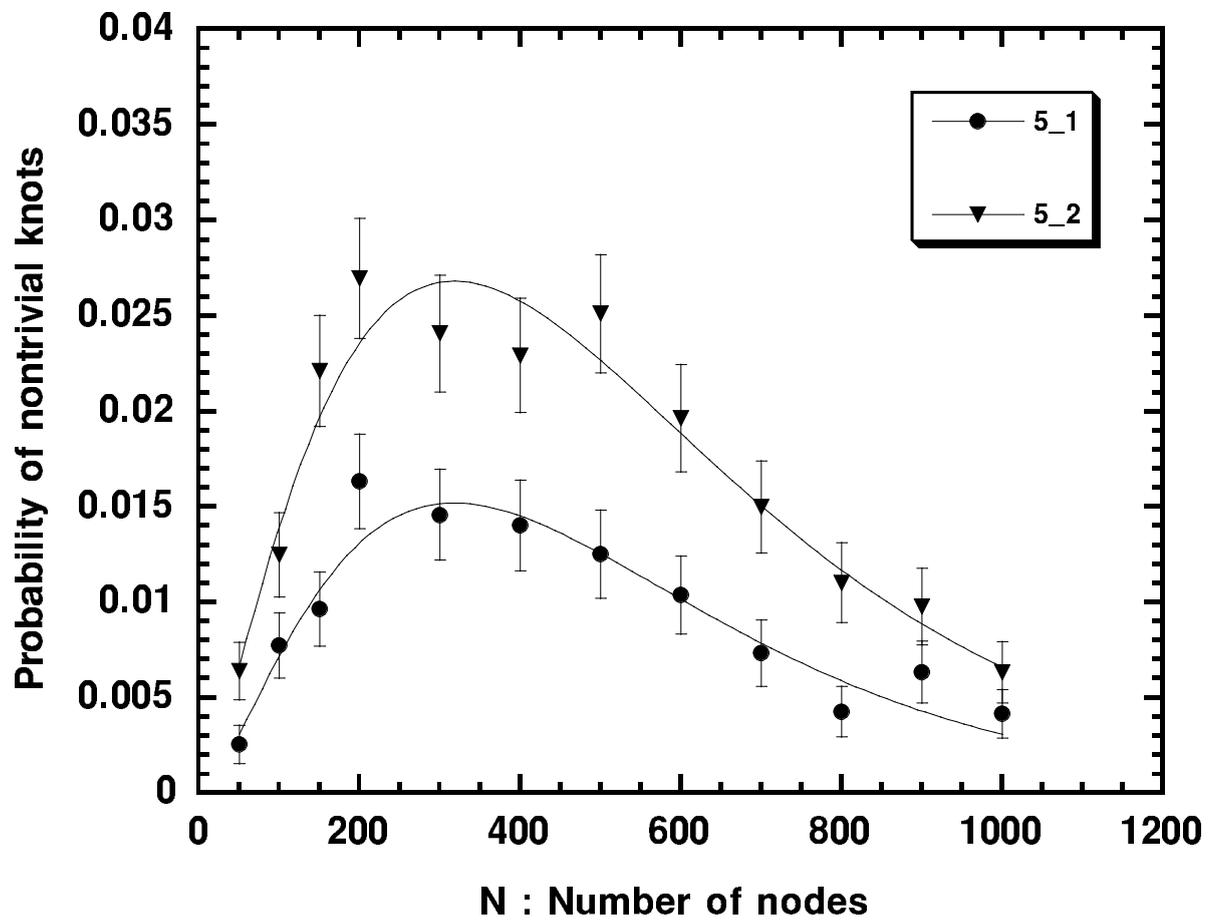
We recall that the knotting probability 
of a nontrivial knot $K$ is given by 
the probability $P_K(N)$ of observing the knot $K$ 
in a given random polygon (or self-avoiding polygon) with $N$ nodes.
Knotting probabilities of several nontrivial knots have been evaluated 
for some different models of random polygons or self-avoiding polygons.
\cite{Vologodskii1,Klenin,DeguchiJKTR,Deguchi95,DeguchiRevE}
Through the simulations 
 with the Vassiliev-type knot invariants \cite{DeguchiJKTR,DeguchiRevE}, 
it has been found that, for the Gaussian random polygons 
and the rod-bead self-avoiding polygons, 
the probabilities of nontrivial knots versus the number  $N$ of nodes 
are well approximated by the fitting curves given by the following formula:   
\begin{equation}
\label{deg}
P_K(N)=C_K \left( N/N_K \right)^{m_K} \exp({-N/N_K}) .
\end{equation}
Here $C_K$, $N_K$ and  $m_K$ are  fitting parameters  to be determined 
from the estimates of the knotting probabilities obtained numerically. 
For a given knot $K$, we call $m_K$ the exponent of the knot.  We also 
call  $N_K$ the characteristic length of the knot $K$.

There are some universal properties 
of the fitting parameters  $N_K$ and  $m_K$ 
of the formula (\ref{deg}). 
For any one of the models investigated, it is found that 
the values of the parameters $N_K$ are almost the same 
for any knot $K$: $N_K \simeq N_0$. \cite{DeguchiJKTR,DeguchiRevE}
Furthermore,  it is observed that the parameter $m_K$ of a knot $K$ 
should be  universal for the different models, 
while  the fitting parameters $C_K$ and $N_K$ are model-dependent. 
\cite{DeguchiRevE} Thus, we may consider that the parameter 
$m_K$ should play a similar role with the critical exponents 
of critical phenomena. 
In fact, the formula (\ref{deg}) of knotting probability 
is  quite consistent with the asymptotic 
scaling behaviors of the number $W(N)$ of all the 
configurations of self-avoiding polygons:
\begin{equation} 
W(N) \simeq C \mu^N N^{\alpha -2} \, , \quad {\rm for} \quad N \gg 1 \, . 
\label{asymp}
\end{equation} 
Here $\mu$ is the growth constant and $\alpha$ corresponds to 
 the critical exponent related to the energy of the $n$-vector model 
through the limit of sending $n$ to 0. 
\cite{deGennes} We can also expect a similar asymptotic expansion 
for the number $W_K(N)$ of a knot $K$.  
Thus, we may call eq. (\ref{deg}) a scaling formula 
of the knotting probability for the knot type $K$.

Let us now consider effects of excluded volume 
on the knotting probability. Through the numerical simulations of 
the cylindrical self-avoiding polygons, 
it has been found that the excluded-volume parameter 
such as the cylinder radius  plays a central role 
in the probability of knot formation.
\cite{LeBret,Klenin}. 
Here, we recall that the probability $P_{knotted}(N, r)$ 
of a cylindrical self-avoiding polygon being a nontrivial knot  
depends on the number $N$ of polygonal nodes,  
the cylindrical radius $r$, and the  Kuhn statistical length  $b$. 	 
It was discussed that the probability  $P_{knotted}(N, r)$ 
should depend on the radius $r$ through  the exponential factor 
$\exp(-\gamma r/b)$ such as shown in eq. (\ref{rb}), where  
 $\gamma$ is the proportionality constant determined from the simulations.
\cite{LeBret,Klenin}.

The empirical formula (\ref{rb}) has been fundamental 
in the study of random knotting of small circular DNAs. \cite{Rybenkov,Shaw} 
However, it is not valid when $N$ is large enough, as we shall see in \S 4.3. 
Here, we  recall that  $N$ is the number of nodes  
of cylindrical self-avoiding polygons. 
Furthermore,  we show in \S 4.4 that our generalized scaling-type  formula 
for the knotting probability of the cylindrical self-avoiding polygons
generalizes  the formula (\ref{rb}).

\subsection{Generalized scaling formula 
for knotting probability of ring polymers}

Let us discuss  the knotting probability $P_K(N,r)$ of the model 
of cylindrical self-avoiding polygons constructed by the dimerization method
that depends not only on  the number $N$ of nodes  but also  on the  cylinder radius $r$.
Generalizing the scaling formula (\ref{deg}) which explains only the $N$-dependence, 
we introduce the following formula  
for the probability $P_K(N,r)$:
\begin{equation}
\label{miyuki}
P_K(N,r)=C_K(r) \left( N/N_K(r) \right)^{m_K(r)} \exp({-N/N_K(r)}) . 
\end{equation}
Here $C_K(r)$, $m_K(r)$, $N_K(r)$ are fitting parameters.
As we shall see later in this section, 
eq. (\ref{miyuki}) gives good fitting curves to numerical estimates 
of the knotting probability of the cylindrical self-avoiding polygons constructed by the dimerization method.

The formula (\ref{miyuki}) also generalizes 
another formula of the knotting probability. 
In a previous paper \cite{PLA}, we have discussed how the probability 
of unknot should depend on the cylinder radius $r$ for the self-avoiding 
polygons consisting of cylindrical segments with radius $r$. 
We note that  the cylindrical self-avoiding polygons in Ref. \cite{PLA} 
are constructed  by the same method discussed in \S 2. 
Through numerical simulation, we have found that 
the unknotting probability denoted by $P_0(N,r)$ 
 can be well described by the  following:
\begin{equation}
\label{pretri}
P_0(N,r)=C_0(r) \exp{(-N/N_0(r))} \, , 
\end{equation}
where $C_0(r)$, $N_0(r)$ are fitting parameters.
We note that 
this formula (\ref{pretri}) corresponds to 
the special case of the general formula (\ref{miyuki})
when $K=0$ and $m_K(r)=0$.
We note  that the radius-dependence of  
the characteristic length $N_0(r)$ of unknot can be approximated by 
an exponential function of the cylinder radius $r$ \cite{PLA}:
\begin{equation} 
N_0(r) \approx N_0(0) \exp(\alpha r ) \, . 
\end{equation} 
We also note that the exponential dependence is favorable to 
the standard theory of polymers. \cite{Nechaev}

\subsection{Numerical estimates of knotting probability
for the model of cylindrical self-avoiding polygons}

Let us discuss the numerical data of the knotting probability 
obtained in our numerical simulations. 
 For a given number  $N$ of nodes, we construct $M$ cylindrical 
self-avoiding polygons with cylinder 
radius $r$ by the dimerization method. 
Here $M$ is given by  $M=10^4$,  
 the polygonal nodes $N$ from 50 to 1000 by 50 or 100, and 
 the radius $r$ from 0.0 to 0.007 by 0.001.
(For details, see Figure Captions.)
We have obtained numerical estimates of 
the knotting probabilities for the trivial knot,  
four prime knots ($3_1$,$4_1$,$5_1$,$5_2$), and two composite knots 
($3_1 \sharp 3_1$,$3_1 \sharp 3_1\sharp 3_1$).
We note that we shall also discuss the knotting probabilities of small sized 
self-avoiding polygons such as $N=20$ or 30 by the cylindrical ring-dimerization method.
In \S 4, the connection to the hedge-hog method is studied. 

In Fig. 4, the estimates of the knotting probability $P_0(N,r)$ 
for the trivial knot are plotted against the number $N$ of nodes  
with $r$=0.001, 0.003, 0.005.
The lines in Fig. 4 are theoretical curves given by the formula (\ref{pretri}).
We confirm the exponential decay of $P_0(N,r)$ also for 
the cylindrical self-avoiding polygons with large numbers of nodes 
which are constructed by the dimerization method. 
(See also Ref.\cite{PLA}.)

In Fig. 5, the numerical values of the knotting probabilities $P_K(N, r)$ 
are plotted against the step number $N$ for the  
prime knots $3_1$, $4_1$, $5_1$. 
We see from  Fig. 5 that  the majority of nontrivial prime knots are given by 
the trefoil knots  ($3_1$).  We also observe that 
the peak values of the  knotting probabilities 
decrease as the cylinder radius $r$ increases. 
There is also a tendency for the prime knots that 
 knots with larger crossing number have less peak values.

In Fig. 6, we show our estimates 
of the knotting probabilities of knots $5_1$ and $5_2$ 
versus the number $N$ of nodes. We see 
that the knotting probability of the knot $5_2$ is always larger 
than that of the knot $5_1$. However, 
for both the two knots 
the knotting probabilities have their 
maximal values at  almost the same number $N$ of nodes.   
Here we note that the error bars of Fig. 6 correspond 
to half of the values 
of the standard deviations of the knotting probability: 
In Figs. 4, 5 and 7, the error bars present  
the standard deviations of the knotting probability,  
while in Fig. 6 they show half of the values since they 
 are too large to be depicted. 
We discuss how we have evaluated the ``standard deviations''
later in \S 3.3.

In Fig. 7, the knotting probabilities of  two composite knots 
$3_1 \sharp 3_1$ and $3_1 \sharp 3_1\sharp 3_1$, 
are plotted against the number $N$ of nodes 
for the values of the cylinder  radius $r$ given by 
$r$=0.001, 0.003, 0.005.
The knotting probabilities of the two composite knots 
have their maxima at different values of the 
number $N$ of nodes, which are also 
different from that of the prime knots.

In  Figs. 5-7 of the knotting probabilities, all 
the fitting curves are given by the formula (\ref{miyuki}). 
 From the Figures, 
we see that it gives good fitting curves 
to the graphs of the knotting probability $P_K(N,r)$ versus 
the number $N$ of nodes for the cylindrical self-avoiding polygons
 with the different values of the radius $r$.
The least-square estimates of the parameters $m_K(r)$, $C_K(r)$, and $N_K(r)$ 
for the fitting curves are listed in Table \ref{para} together 
with their $\chi^2$ values.
The errors for the best estimates of the  parameters given in Table \ref{para}
 correspond to 68.3 $\%$ confidence intervals, which are 
equivalent to the standard deviations.

Let us explain the method for evaluating  
the errors of the estimates of the knotting probabilities:   
with the error values,  we have determined 
 the fitting curves shown in Figs. 4-7.   
Throughout the paper,  we  estimate  
the error of the knotting probability $P_K(N,r)$ 
by the following method: taking the sum of the contributions from 
the statistical fluctuation of the number $M_K$ 
and that of statistical weight appearing in the ring-concatenating procedure 
(i.e., the ring-dimerization process in \S 2.3), 
 multiplying the sum  by the factor of 2, 
and then we regard the result as the error corresponding to  
the standard deviation of the knotting probability $P_K(N,r)$. 
It seems that 
this  method of evaluating errors  might give  larger values of errors 
than other methods. However, we have employed it 
in order to compensate some possibly neglected contributions to 
the errors 
arising from the chain-dimerization process. It    
 would  be not easy  to estimate the possible errors 
 in the dimerization process, systematically.  
In fact, we do not take into account 
any possible statistical fluctuation  in the dimerization process 
 for constructing  cylindrical self-avoiding chains.     
It could be as large as that of 
 the statistical weight 
in the ring-concatenating procedure.

We make a comment on the dimerization process.  
Giving  one initial number to the generator of pseudo-random numbers, 
we construct $M=10^4$ cylindrical self-avoiding polygons 
with number $N$ of nodes and radius $r$.  
This method makes our simulation simpler and more practical than 
giving several different initial numbers to the generator. 
However, it might give  slightly larger values of errors.    
This consideration has  been reflected in our method 
for estimating errors  discussed in the last paragraph.

\subsection{Some novel properties  of fitting parameters}

Let us discuss the best estimates of the fitting 
parameters $N_K(r)$, $m_K(r)$ and $C_K(r)$  listed in  
Table \ref{para}.  From the list, 
we can derive  a number of  important observations 
for the fitting parameters.

First, we consider   the values of the parameter $C_K(r)$. 
 From Table \ref{para}, we see that the values of $C_K(r)$ 
are almost independent of the cylinder radius $r$, within the error bars.  
Thus we may have the conjecture that the parameter 
$C_K(r)$ should be  a constant with respect to the radius $r$ for any knot $K$:   
\begin{equation}  
C_K(r)  \simeq C_K(0) .
\end{equation}

The parameter $C_K(0)$ of a knot $K$ must reflect at least some kind of 
 complexity of the knot. In fact, 
the probability of observing the knot type $K$ is small
 when the value of $C_K(0)$ is small.
 Furthermore, as far as the prime knots are concerned, 
 there seems to be a tendency that 
 a knot with a smaller value of the knotting probability    
 is likely to be more complex.  
Furthermore, 
 the value of the parameter $C_K(0)$ 
is not directly related to the crossing number of 
the knot $K$. 
For instance,  the value of $C_{5_1}(0)$ 
is smaller than that of $C_{5_2}(0)$ 
although both of the knots  $5_1$ and $5_2$ have the same crossing number.
  
Second, we consider the characteristic length $N_K(r)$. 
Let us  discuss how  the parameter  $N_K(r)$ should depend on 
the cylinder radius $r$ for a given knot type $K$.  
 From Fig. 8, we see that  
 $N_K(r)$ can be approximated 
by an exponential function of the chain thickness $r$ :
$N_K(r)= N_K(0) \exp(\alpha_K r)$, at least    
for the trivial knot $K=0$ and the trefoil knot $K=3_1$. 
Considering  the poor statistics for the case of the nontrivial knots, 
we may conjecture  that $N_K(r)$ should  depend on $r$ exponentially for 
any knot $K$. 
Let us next discuss the knot-dependence of the parameter $N_K(r)$ 
with the radius $r$ fixed. From Fig. 9, we see that 
the values of the parameters $N_K(r)$ for the different knots are almost 
given by the same value  with respect to  the errors.

 Combining  the results of 
the radius- and knot- dependence of the parameter $N_K(r)$, we have 
 the conjecture that the parameter $N_K(r)$ for a knot $K$ should be equal to 
that of the trivial knot and also that 
it should be given by an exponential function of 
the cylinder radius $r$:  $N_K(r) = N_0(0) \exp( \alpha_0 r)$.

In fact, in the previous paper \cite{PLA}, 
 we calculated the probability of observing the trivial knot 
from the values of the cylinder radius 
$r$  from 0.0 to $0.1$ by 0.01, 
and discussed how the parameter $N_0(r)$ 
should depend on the cylinder radius $r$. Then, 
we found that the characteristic length $N_0(r)$ is roughly approximated by an increasing exponential function of $r$ as $N_0(r)=N_0(0) \exp({\alpha r})$, with $N_0(0)=292 \pm5$, and $\alpha=43.5 \pm 0.6$. We note that $\chi^2$=42 while the number 
of data points is ten with the two fitting parameters. 
In order to improve the $\chi^2$ value, 
we have also considered another fitting formula $N_0(r)=N_0(0)\exp({\beta r^\nu})$, 
where there are three parameters to fit $N_0(0)$, $\beta$, and $\nu$.
The best estimates of the fitting parameters are given by $N_0(0)$=271$\pm$6, 
$\beta=29\pm2$, $\nu=0.85\pm0.02$, and  the $\chi^2$ value is given by 2.2 .

The results of the previous paper \cite{PLA} 
are consistent with that of the trivial knot in this paper.
We note that the range of the cylinder radius $r$ in the present paper 
is much narrower than 
that of Ref. \cite{PLA}. Here, we consider 
the values of the cylinder radius from 0.001 to  0.007 by 0.001.  
Even the second fitting function: $N_0(r)=N_0(0)\exp({\beta r^\nu})$ 
can also be approximated by an exponential function of $r$. 
However, the connection between the formulas (\ref{miyuki}) and (\ref{rb})(or 
(\ref{volo}))
can be shown more clearly by the first fitting formula than by the second one as we shall show in \S 4.
Thus, we employ the first one in the paper.

Third, we  discuss the exponent $m_K(r)$ of a knot $K$.
 From Table \ref{para}, we see that the values of the parameter  $m_K(r)$ 
should be independent of the cylinder radius $r$ for any knot $K$. 
This  property is consistent with the conjecture that the exponent $m_K$ of 
a knot $K$ should be universal, which was also investigated for  
the rod-bead model of self-avoiding polygons and 
the Gaussian model of random polygons. \cite{DeguchiRevE}  
Furthermore, within the error bars,  
the values of the exponents can be roughly approximated as follows: 
 $m_K(r) \simeq 0$ for the trivial knot ($K=0$);  
 $m_K(r) \simeq  1$ for the four prime knots $3_1$, $4_1$, $5_1$ and $5_2$; 
$m_K(r) \simeq 2$ for the composite knot $3_1 \sharp 3_1$;  
$m_K(r) \simeq 3$ for the composite knot $3_1 \sharp 3_1 \sharp 3_1$. 
It is interesting to note that the roughly approximated values of $m_K(r)$ 
are consistent with the results not only of the off-lattice models 
such as  the Gaussian model \cite{DeguchiJKTR}
and the rod-bead model \cite{Deguchi95}, but also  of the lattice model \cite{Orlandini} 
of self-avoiding polygons. The roughly approximated values are also 
consistent with the additivity of the exponent $m_K$ that the exponent of a composite knot 
should be  given by the sum of the exponents of the constituent prime knots:  
$m(K_1 \sharp K_2) = m(K_1)+m(K_2)$ and so on,   
which was first observed 
for the Gaussian model of random polygons. \cite{DeguchiJKTR}

Let us discuss an interesting consequence of the conjectures 
on the fitting parameters given in the above.   
We first note that 
among the three fitting parameters, only the characteristic length 
$N_K(r)$ should depend on the cylinder radius $r$;
 the other  two parameters should be independent of it.
We recall also the conjecture that $N_K(r)=N_0(r)$ for any knot $K$. 
Thus, if we assume that 
the value of the  exponent $m_K$ 
can be approximated  by the same value for any prime knot $K$, 
then we  have the following relation between the knotting probabilities 
of any two prime knots $K_1$ and $K_2$:
\begin{equation} 
P_{K_1}(N,r)/P_{K_2}(N,r) \simeq C_{K_1}/C_{K_2}  . 
\end{equation} 
Thus, the  ratio of the knotting probabilities of any two prime knots 
$K_1$ and $K_2$ 
should be roughly 
given by  that of the parameters $C_{K_1}$ and $C_{K_2}$.

\setcounter{equation}{0} 
\renewcommand{\theequation}{4.\arabic{equation}}
\setcounter{section}{3} 

\section{Topological Entropy and Random Knotting of Circular DNAs}

\subsection{Probability of random knotting of circular DNAs}

In this section, we  discuss the connection of the knotting probability 
to some experiments of  DNA random knots \cite{Rybenkov,Shaw}.

Let us consider  DNA molecules in an electrolyte solution. 
The DNAs are polyelectrolytes with negative charges,  and 
the DNA chains are surrounded by some clouds of counter ions.   
Partially due to the electrostatic repulsion among the chains, 
they are considered as stiff chains; in fact,  the  Kuhn statistical length 
of DNA chains is given by rather  a large value, about 50 nm 
\cite{Grosberg-book}.  

In order to study thermodynamic properties of DNA chains  
we can simulate  the DNA chains   
 by some configurations of wormlike chains  with the effective diameter 
corresponding to the screening length \cite{Stiger,OSF}. 
For any wormlike chain, the persistent length (i.e., the Kuhn length) 
is fundamental.  
The length of 
a wormlike chain should be expressed in terms of 
the Kuhn statistical unit.   
Thus, we can replace wormlike chains by such self-avoiding walks 
that have cylindrical segments with the persistent length. 
Let us now consider wormlike rings 
which are given by  rings made of wormlike chains. Then,  
we can also approximate wormlike rings by 
such self-avoiding polygons consisting of cylindrical segments 
with some effective diameter 
whose length is given by the Kuhn length \cite{Klenin}.

Let us discuss the algorithm for generating
cylindrical self-avoiding polygons  
 which we call the hedge-hog method of ring polymers.
 \cite{Klenin} 
The algorithm is given by  the following: 
we  generate a set of vectors of unit length 
with a common origin (a ``hedge-hog''),  
and applying the Monte-Carlo procedure to the set, we derive 
a random sequence of self-avoiding polygons, 
keeping only those configurations that have no overlap  
between any two unadjacent cylindrical segments. 
(See \S  A.2)

Through the numerical simulations, 
 the probability of being knotted has been evaluated for the 
hedge-hog method \cite{Klenin}.
 For the data, 
the radius-dependence of the probability 
$P_{knotted}(N,r)$ has been well approximated 
 by the empirical formula (\ref{rb}).  
Furthermore, a different method of wormlike polygons 
 has been introduced, which we call the MC method with the bending energy
(See \S A.3) \cite{Rybenkov},  and  
the knotting probabilities of 
two prime knots $3_1$ and $4_1$ 
are evaluated 
for the wormlike rings
with $N$ Kuhn lengths, where   
$N$ is  given from 16 to 60.   
 For a given nontrivial knot $K$,   
let the symbol  $P_K(N,r)$  denote 
the knotting probability 
of the wormlike rings, which are constructed by the MC method 
with the bending energy
with the chain radius $r$ 
consisting of $N$ units of the Kuhn length $b$. 
 Then, the data are fitted by   
the following formula: 
\begin{equation}
\label{volo}
P_K(N,r)=P_K(N,0) \exp({- \gamma_K \frac{r}{b}})
\end{equation}
Here, $\gamma_K$ 
is a constant which is independent of $N$ and to be determined 
by the simulation for each nontrivial knot $K$ \cite{Rybenkov}.

 The knotting probability of the wormlike rings 
of the MC method with the bending energy 
is consistent with  that of the hedge-hog method; 
the knotting probability of  
 wormlike rings of the MC method with the bending energy
 consisting of $N$ Kuhn units with 
the cylinder radius $r$ 
should be equivalent to  the hedge-hog method with 
$N$ cylindrical segments with radius $r$.
In fact, it was discussed in Ref. \cite{Rybenkov} 
that the  knotting probability of  a particular knot
for the wormlike ring can be determined by the number $N$ of 
Kuhn statistical units and the chain radius $r$.  
In general, for any model of wormlike rings,
the Kuhn statistical length $b$  
is not necessarily equivalent to the length of one  segment of the rings; 
it can be much longer than the length of the segments,
such as the case of the MC method with the bending energy.

The empirical formulae (\ref{rb}) and (\ref{volo}) give 
the connection of the knotting probability 
to experiments of DNA random knots. 
V.V. Rybencov $et. al.$ and S. Y. Shaw and J.C. Wang 
measured the fractions of knotted species generated 
by the method of random ring closure of DNAs 
through the electrophoretic separation method.  \cite{Rybenkov,Shaw}
The effective diameter of the DNA double helix was evaluated, 
by comparing the knotted fractions observed in the experiments 
with the theoretical estimates of the 
knotting probability given by the computer simulations \cite{Klenin}   
of the hedge-hog method. \cite{Rybenkov,Shaw}

\subsection{Consistency of knotting probabilities for
 dimerization and hedge-hog methods}

We now  discuss that the hedge-hog and dimerization methods for constructing
cylindrical self-avoiding polygons are consistent as far 
as their knotting probabilities are concerned. 
Let us consider our numerical estimates of 
the  probability $P_{knotted}(N,r)$ of being knotted 
for the  hedge-hog and dimerization methods, which are obtained in our 
simulations.   Here, we set $b=1$ for the hedge-hog method without any loss of generality.

 For the hedge-hog method, 
we have evaluated the probability $P_{knotted}(N,r)$ 
for the cases of  $N$=20 and 30.  
 For the dimerization method, we have evaluated it  for the cases of
 $N$=21 and 31.   
 For each of the methods, the values of the diameter $2r$ are  
  given by the values from 0.0 to 0.05 by 0.01.  
 
 Our numerical estimates of the probability  $P_{knotted}(N,r)$ 
 of the hedge-hog method gives almost the same with 
 that of Ref. \cite{Klenin}.  Furthermore, our estimates of the 
  probability $P_{knotted}(N,r)$ 
 of the hedge-hog method is also consistent with 
 that of the dimerization method.
 We have thus confirmed  that the data of the probability  $P_{knotted}(N,r)$ 
 of the cylindrical self-avoiding polygons of the dimerization method
 satisfy  the empirical formula (\ref{rb}) for 
 the thickness-dependence of the knotting probability, 
for the small numbers $N$ of nodes.

\subsection{Limited validity of the empirical formula 
 for the  knotting probability of DNAs }        

We recall that  
the empirical formula (\ref{volo}) describing 
the thickness-dependence of the knotting probability has been fundamental 
in the study of the random knots of circular DNAs. However, we shall show that 
 it is not valid when the number $N$ of nodes is  large enough.

Let us consider  the $N$-dependence of the ratio: $P_K(N,r)/P_K(N,0)$ of 
the knotting probability for a given nontrivial knot $K$. 
If the empirical formula (\ref{volo}) should be valid, 
then the ratio would be given by the following: 
\begin{equation} 
P_K(N,r)/P_K(N,0) = \exp(-\gamma_K r/b) \, .  
\end{equation}  
We note that it should be  constant with respect to $N$.

In Fig. 11, the data of $P_K(N,r)/P_K(N,0)$ 
for the region of $r$  from $r$=0.001 to $r$=0.007 by 0.001 
are plotted against the number $N$ of nodes.
Then, we see that the ratio is not constant 
but increases with respect to $N$. 
Thus, the formula (\ref{volo}) does not hold when the number $N$ of nodes 
is large enough such as  $N > 100$.

We make a comment on the method for estimating the errors shown in Fig. 11. 
We note that  the error bars denote the standard deviations in Fig. 11. 
The variance $\sigma^2$ of the ratio: $P(N,r)/P(N,0)$ 
is given by the following formula: 
\begin{equation}
\sigma^2=\frac{1}{P(N,0)^2} \left\{ \sigma_1^2 
+ \left(\frac{P(N,r)}{P(N,0)}\right)^2 \sigma_2^2 \right\}
\end{equation}
Here $\sigma_1$ denotes the variance 
of $P(N,r)$ and $\sigma_2$  that of $P(N,0)$.

\subsection{Reduction  of the scaling formula 
of knotting probability in the case of small $N$}

Let us explicitly discuss how the generalized scaling formula (\ref{miyuki}) 
is related to the empirical formula (\ref{volo}). 
In fact, this gives a kind of ``extrapolation"  
of the scaling formula into a region of small $N$ .   
The scaling formula should be valid when $N$ is very large, 
since it is nothing but an asymptotic expansion with respect to $N$.  
On the other hand,  the empirical formulae (\ref{rb}) and (\ref{volo}) 
should be  valid only for some small number $N$ of nodes.

There are three points in our discussion. 
First, we recall that the parameter $C_K(r)$ should be independent 
of the cylinder radius $r$: $C_K(r) \simeq C_K(0)$.  Then, 
from the formula (\ref{miyuki}), we have the following expression 
for the ratio of $P_K(N,r)$ to $P_K(N,0)$:  
\begin{equation}
 \frac{P_K(N,r)}{P_K(N,0)}= \left( \frac{N_K(0)}{N_K(r)}\right)^{m(K)} 
\exp\left( {\frac{N}{N_0(0)}} - {\frac{N}{N_0(r)}}  
\right) 
\label{ratio} 
\end{equation}
Second,  we can neglect the 
exponential factor in the right-hand-side of (\ref{ratio}), when $N$ is small such as $N \le 30$.
The characteristic length $N_K(r)$ can be  roughly evaluated 
 as $N_K(r) \ge 300$ when  $r$ is given by some value from 0.001 to 0.01. 
The value is about ten times larger than that of 
the number $N$ of nodes, when $N < 30$.   
Therefore, the exponential argument
$\frac{N}{N_0(0)}-\frac{N}{N_0(r)}$ should be  rather small. 
Third, we  recall 
 that the characteristic length $N_0(r)$ can be given by 
 $N_0(r) \simeq  N_0(0) \exp({\alpha r})$, 
as discussed  in \S 3.4  (see also in the previous paper: \cite{PLA}).    
Furthermore,  we may assume the conjecture 
that the characteristic lengths do not depend on any knot type 
$K$: $N_K(r) \simeq N_0(r)$. 
Combining the three points given in the above, we see that   
 the main contribution to the ratio of eq. (\ref{ratio}) is given by 
\begin{equation}
 \frac{P_K(N,r)}{P_K(N,0)} \simeq 
\exp\left( {-\alpha m_K r} \right)
\label{exponent} 
\end{equation}
We note that 
 the factor $(N_K(0)/N_K(r))^{m_K}$ should correspond  to 
the exponential factor  $\exp(- \gamma_K r/{b})$ 
in the empirical formulae (\ref{rb}) and (\ref{volo}). Explicitly, we have 
$\gamma_K = \alpha m_K$ when $b=1$.  
Thus it is shown that the empirical formula (\ref{rb}) can be derived from the generalized scaling formula (\ref{miyuki}).
We recall that the constant $\gamma_K$ is assumed to be independent of $N$ in Ref. \cite{Rybenkov}.

We remark that the knotting probability $P_K(N,r)$ can be described
by the generalized scaling formula (\ref{miyuki}) from small $N$ to large $N$,
as shown in \S 3.3, explicitly.
Furthermore, if we assume the properties $C_K(r)\simeq C_K(0)$,
$N_K(r)\simeq N_0(r)$ and $N_0(r)\simeq N_0(0) \exp (\alpha r)$,
then we have the following expression
\begin{equation}
\log \left( \frac{P_K(N,r)}{P_K(N,0)} \right)
\simeq -\alpha m_K r +\frac{N}{N_0(0)}(1 -\exp (-\alpha r)),
\end{equation}
which is consistent with the $N$-dependence shown in Fig. 11, 
within error bars.

For the case of the trivial knot, we may also consider the 
$r$-dependence of the second factor of the right-hand-side of eq. 
(\ref{ratio}),  
which can be neglected for the case of nontrivial knots. 
Taking into account the fact that $m_0 \simeq 0$, 
we have the following approximation:
\begin{equation} 
 \frac{P_0(N,r)}{P_0(N,0)} \simeq 
\exp\left( {\frac N {N_0(0)}}(1- \exp(- \alpha r)) \right) \, . 
\end{equation} 
Here we note that the $r$-dependence of the unknotting probability 
has not been discussed previously, yet.

\setcounter{equation}{0} 
\renewcommand{\theequation}{5.\arabic{equation}}

\section{ Concluding Remarks}

In this paper we have discussed 
how the knotting probability (or equivalently, the topological entropy) 
should depend on the cylinder radius 
for the cylindrical self-avoiding polygons.
The dimerization and hedge-hog methods give almost the same values for 
 the knotting probability, although their algorithms  are quite different.
This coincidence suggests that 
any algorithm of cylindrical self-avoiding polygons 
with $N$ Kuhn statistical units  and the cylinder radius $r$ 
may give essentially the same value for the knotting probability.   
We have also found that for any  knot investigated,  
the knotting probability of the model of cylindrical self-avoiding polygons
 is  described by 
the generalized scaling formula (\ref{miyuki}) 
as a function of the number $N$ of nodes and the cylinder radius $r$.

 From the best estimates of the fitting parameters, 
we have observed several important properties of the parameters. 
Based on  these properties,
we can derive a conjecture of the best formula of the knotting probability. 
For   self-avoiding polygons with 
$N$ cylindrical segments with the radius $r$  of unit length,  
the best formula of the knotting probability for knot $K$ is given by the following:   
\begin{equation} 
P_K(N,r) = C_K \left(N / N_c(r) \right)^{m_K} \exp(-N/N_c(r)) \,  ,
\end{equation} 
where the characteristic length $N_c(r)$ is independent of knots  
and is given by 
\begin{equation} 
N_c(r) = N_c(0) \exp( \alpha r) \, . 
\end{equation}
We recall that $C_K$ and $m_K$ should be  constant with respect to 
the cylinder radius $r$. 
We shall investigate the conjecture 
more precisely in later publications.

 Finally, we make a comment on the range of $r$.
In the paper, we have discussed for some nontrivial knots the knotting probabilities of the cylindrical self-avoiding polygons with radius $r$ given from 0 to 0.007.
On the other hand, in Ref. \cite{PLA}, we have discussed the probability of unknot for the wider range of $r$:
from 0 to 0.1.
From the result of Ref. \cite{PLA}, it is suggested that the conjecture (5.1)
should be valid also for the larger values of radius $r$.
Thus, it is an interesting future problem to check the conjecture explicitly through numerical simulations.

\setcounter{equation}{0}
\renewcommand{\theequation}{A.\arabic{equation}}
\setcounter{section}{0}
\renewcommand{\thesubsection}{A.\arabic{subsection}}
{\bf Appendix A: Algorithms for the Model of Ring Polymers}

In Appendix A, we describe the three different algorithms for
constructing  self-avoiding polygons consisting of freely jointed hard
cylinders.\cite{PLA,Klenin,Rybenkov}
All the three algorithms should produce equivalent sets of
ring polymers corresponding to the discrete worm-like chains,
if the  numbers of polygons constructed are very large.
As far as constructing large polygons,  however, it seems that
the dimerization algorithm should be the most efficient.

\subsection{Cylindrical ring-dimerization method (the dimerization
method)}
Let us explicitly discuss the method
for  generating ring polymers of freely jointed hard cylinders, by which
almost all the polygons in the paper are constructed.
The method is based on the algorithm of Ref. \cite{Chen} for the rod-bead
model.
It consists of three  parts:
(1) generation of basic chains of hard cylinders; (2) propagation of
linear chains by the dimerization algorithm;
(3) formation of ring polymers by connecting
two linear chains. We call the combined procedure of the parts (1), (2) and
(3), the cylindrical ring-dimerization method.

\par
Basic chains consisting of hard cylinders
are generated by the straightforward Monte Carlo method.
For an illustration, let us suppose that we make basic chains
of eight cylindrical segments.
Then, for any pair of unadjacent cylinders, we check
the overlapping condition between the  cylinders which is described in Appendix B.
If there is a pair satisfying the overlapping condition in a given basic
chain,
then we throw away the whole chain and start from the beginning.
 If we find that there is no overlapping pair of unadjacent cylinders,
then we store the chain in a computer disk.
Here we recall that  we do not check the overlapping condition for
any  pair of adjacent cylinders. This  corresponds to the freely-jointedness
of the model.

\par
Let us consider how to make chains of sixteen cylinders.
We make  two pools of $M$ basic chains in the disk. Then, we choose randomly
basic chains 1 and 2   from the first and second pools, respectively.
We  join them together by placing the zeroth node of the second
chain on the eighth node of the first chain. If there is no overlapping pair
of
unadjacent cylinders in the new chain of sixteen cylinders,
then we store it in the computer disk.  If there is any overlapping
unadjacent pair, then we throw it away. We thus construct $M$ chains of
sixteen cylinders
in the disk. The same procedure is used for
creating  chains of , say, 32 cylinders, 64 cylinders, etc.
We note that
 chains with any number of cylinders can be constructed
by the dimerization scheme shown in Fig. 3.

\par
Let us describe
the procedure of making ring polymers of $2N+1$ cylinders
from linear chains of $N$ cylinders. We consider two pools $S_1$ and $S_2$
of
$M$ chains of freely jointed $N$ hard cylinders.
We pick up randomly chain 1 in  the pool $S_1$.
Then, we can choose  chain 2 randomly in the pool $S_2$.
In order to make the process more efficient, however, we choose chain 2 as
follows.
Let the symbols $h_1$ and $h_2$ denote the end-to-end distances of chains 1
and 2, respectively.
Here we recall that the length of  cylindrical segments is given by 1.
Then, we choose chain 2 randomly
from the group of chains in $S_2$ which satisfy
the condition:
\begin{equation}
|h_1 - 1| \le h_2 \le h_1  + 1 . \label{h12}
\end{equation}
The sampling bias induced by this operation is corrected
with the statistical weight $m/M$, where $m$  is the number of chains in
$S_2$
which satisfy the condition (\ref{h12}). We recall that $M$ is the total
number of
linear chains in $S_2$.

\par
Let us put a new cylindrical segment  between the two ends of the
chains 1 and 2. Then, we check whether there exists any  segment
in the two chains  which overlaps with the new segment, except for the two
neighboring ones.
We also check the overlapping condition for every unadjacent pair of cylinder in which are cylinder is in chain 1 and another in chain 2.
If there is no overlap, then we consider that the selected pair of
chains makes a perfect ring with $2N+1$ cylindrical segments.
We store the ring in the hard disk.
However, the probability of forming a ring depends on the
values of $h_1$ and $h_2$. Let us denote by $\theta$ the angle between
the end-to-end vectors of the two chains 1 and 2. Then, the probability
is proportional to $2\pi \sin \theta$. We note that
$2\pi h_2 \sin \theta $ is the arclength
of the circle on which the end-point of the chain 2 can be placed.
The probability of forming a ring
should be proportional to the arclength of the circle divided by $h_2$.
Thus, the total statistical weight
of the ring-dimerization procedure is given by
\begin{equation}
W = {\frac m M } \sin \theta
\label{weight}
\end{equation}
Here, the angle $\theta$ is determined  by the values $h_1$ and
$h_2$ as follows
\begin{equation}
\cos \theta = {\frac {h_1^2 + h_2^2 - 1} {2 h_1 h_2}  }
\end{equation}
We remark that all the expectation values of some quantities of the model
such as
the knotting probability etc.,
should be calculated by taking the weighted averages with respect to the
statistical weight of eq.  (\ref{weight}).

\subsection{Monte-Carlo method with random hedgehog (the hedge-hog
method)}

Let us describe the algorithm of Ref. \cite{Klenin}, which we call the
Monte-Carlo method
with random hedgehog, or the hedgehog method, for short.
First, a set of $n$ (even number) vectors $\boldmath{e_i}$
of unit length with a common origin (a ``hedgehog'')
 is generated as follows:
 In the hedgehog, for those vectors that have odd $i$ values, their
directions are
chosen randomly and independently;
for even $i$ values, we set  $\boldmath{e_i=-e_{i-1}}$.
Thus, we have $\sum_{i=1}^{n}\boldmath{e_i}=0$.
Then,  to exclude pair correlations between the vectors,
for each $\boldmath{e_i}$, we choose a  different vector $\boldmath{e_j}$
randomly
 from the hedgehog and the pair of vectors $\boldmath{e_i}$
 and $\boldmath{e_j}$ are rotated at a random angle around
the bisectrix of the angle between them.
This operation does not change the sum $\boldmath{e_i+e_j}$, and
the sum of all the  vectors in the hedgehog remains zero, consequently.
The process is repeated for each vector of the random hedgehog, many times.
As a result, a ``random hedgehog'' is obtained.
Then, the vectors of the random hedgehog
are put in an arbitrary order, and we have the chain.
The resulting chain is automatically closed.

\par
In order to take into account the excluded volume effect,
chain segments are modeled as hard cylinders with radius $r$.
We generate a large number of closed chains  by
the random hedgehog method,
and then retained only those of them for which the minimum distance between
any two unadjacent segments is larger than the segment diameter $2r$.
%
%

\subsection{Monte-Carlo method with the bending energy}

We  consider the algorithm of Ref. \cite{Rybenkov}, which we call
the Monte-Carlo method with the bending energy.
A closed chain  of $kn$ segments of rigid impenetrable cylinders
 of equal length and radius $r$ is constructed.
Here $k$ elementary segments correspond to  the Kuhn statistical length,
and the closed chain consists of $n$ Kuhn segments.
The conformational sets are obtained by successive deformations
of a starting conformation in accordance with the Metropolis-Monte Carlo
procedure.
The deformation is rotation of a subchain containing an arbitrary number of
 adjacent segments by a randomly chosen angle, $\phi$, around the straight
line connecting the vertices bounding the subchain.
The value of $\phi$ is  uniformly distributed over an interval ($-\phi_0$ to
$\phi_0$),
where the interval is chosen so that about half of the moves are accepted.
It depends on the energy whether a trial conformation generated is accepted
or not.
The energy of the chain, $E$, is calculated as
\begin{equation}
E=RT\alpha \sum_{i=1}^{kn}\theta_i^2
\end{equation}
where summation is done over all the joints between the elementary segments,
$R$ is the gas constant, $T$ is the absolute temperature,
$\theta_i$ is the angular displacement of the $i$th segment relative to
segment $i-1$,
and $\alpha$ is the bending rigidity constant.
The bending rigidity constant is chosen so that
exactly $k$ elementary segments correspond to the Kuhn statistical segment
length.
For each set of values of $n$ and $r$, a large number of  conformations are
generated.

\appendix
\setcounter{equation}{0} 
\renewcommand{\theequation}{B.\arabic{equation}}
\setcounter{section}{0} 
{\bf Appendix B: Overlapping Condition of the Cylinder Model}
 
We discuss the condition when a given pair of 
cylinder segments with radius $r$ 
has an ``overlap'', explicitly. 
We first recall that the central line segment 
of a cylinder is defined by the line segment 
between the  centers of the upper and lower disks of the cylinder.
We now define the condition of an ``overlap'' as follows:
 given two cylinder segments are said to have an ``overlap''
if and only if the distance between their central line segments 
is less than the cylinder diameter $2r$.  

\par 
Let us formulate the algorithm for the overlapping condition. 
We consider a pair of two cylinders with radius $r$ and unit length 
in three dimensions. We may assume that the end points of their  
central line segments are given by ${\vec b}$ and ${\vec a}+{\vec b}$, 
${\vec d}$ and ${\vec c}+{\vec d}$, respectively. 
Here, the vectors ${\vec a}$ and ${\vec c}$ are unit vectors. 
Then, any point on the central line segments 
can be expressed  by $\vec{X_s}$ or $\vec{X_t}$ given in the following:  
\begin{equation}
\vec{X_s}=\it{s}\vec{a}+\vec{b} \qquad 
\vec{X_t}=\it{t}\vec{c}+\vec{d} 
\end{equation}
where $s$ and $t$ are real parameters satisfying $0 \le s,t \le 1$. 
We also define  the angle parameter $\theta$ by the relation: 
$\vec{a} \cdot \vec{c} = \cos \theta$,  
where we take the branch: $0 \le \theta \le \pi$. 

\par 
Let us consider the two infinite lines 
extending the two central line segments for 
$\vec{X_s}$ and $\vec{X_t}$, respectively.
 Then, the distance between the two infinite lines is 
given by the minimum of the quantity: $D=\mid \vec{X_s}-\vec{X_t} \mid$. 
Denoting the minimum distance by  $D_{min}$, we have   
\begin{equation}
D_{min}^2=-s_m^2-t_m^2+2 s_m t_m \cos \theta + \gamma \, , 
\end{equation}
where the parameters $s_m$ and $t_m$ are given by 
\begin{equation}
    s_m  = \frac{1}{\sin^2 \theta}(\alpha+\beta \cos\theta) \, , \qquad 
    t_m  = \frac{1}{\sin^2 \theta}(\alpha \cos\theta+\beta) \, . 
\end{equation}
Here $\alpha$, $\beta$, and  $\gamma$ have been defined by  
 $\alpha=-\vec{a}\cdot(\vec{b}-\vec{d})$, 
$\beta$=$\vec{c}\cdot(\vec{b}-\vec{d})$, 
and $\gamma$=$(\vec{b}-\vec{d})^2$, respectively. 
We note that  $s_m$ and $t_m$ are such values of the parameters $s$ and $t$ 
that give the minimum $D_{min}$ for $D=\mid \vec{X_s}-\vec{X_t} \mid$.

\par 
Let us now explain the algorithm. 
First we calculate the value of $\cos \theta$.
Here, we suppose that $\cos^2 \theta < 1$,   
and we shall consider the case when $\cos^2 \theta =1$, later.  
If $D_{min}$ is larger than the diameter $2r$, then 
there is no overlap between the two cylinder segments. 
If $D_{min}$ is smaller than $2r$,
 we check the values of $s_m$ and $t_m$.
If 0$\le s_m \le$1 and 0$\le t_m \le$1, there is an overlap; 
otherwise,  we choose the value $d_{min}^2$, which will be given shortly,  
and we calculate $D_{sum}^2=d_{min}^2+D_{min}^2$.
If $D_{sum}$ is larger than $2r$, then there is no overlap; 
otherwise, there is an overlap.

\par 
Let us now formulate the value $d_{min}^2$, explicitly. 
We first define $z_j$ for $j=0,1,2, 3$, in 
the following: 
\begin{eqnarray} 
z_0 & = & s_m - t_m \cos \theta \nonumber \\ 
z_1 & = & t_m - s_m \cos \theta \nonumber \\ 
z_2 & = & z_0 + \cos \theta \nonumber \\ 
z_3 & = & z_1 + \cos \theta  
\end{eqnarray} 
Then, we define $p_j$ for $j=0, 1, 2, 3$ as follows:
\begin{eqnarray} 
p_0 & = & \left\{
\begin{array}{@{\,}ll} 
z_0^2 + t_m^2 \sin^2 \theta  & {\rm for} \quad z_0 < 0  \\ 
 t_m^2 \sin^2 \theta  &  {\rm for} \quad 0 \le z_0 \le 1  \\ 
 (1-z_0)^2 + t_m^2 \sin^2 \theta  &   {\rm for} \quad z_0 > 1    
\end{array}
\right. 
\nonumber \\
p_1 & = & \left\{
\begin{array}{@{\,}ll} 
z_1^2 + s_m^2 \sin^2 \theta  &   {\rm for} \quad z_1 < 0  \\ 
 s_m^2 \sin^2 \theta  &   {\rm for} \quad   0 \le z_1 \le 1  \\ 
 (1-z_1)^2 + s_m^2 \sin^2 \theta  &   {\rm for} \quad  z_1 > 1    
\end{array}
\right. 
\nonumber \\
p_2 & = & \left\{
\begin{array}{@{\,}ll} 
z_2^2 + (1-t_m)^2 \sin^2 \theta  &   {\rm for} \quad z_2 < 0  \\ 
 (1-t_m)^2 \sin^2 \theta  &   {\rm for} \quad 0 \le z_2 \le 1  \\ 
 (1-z_2)^2 + (1-t_m)^2 \sin^2 \theta  &   {\rm for} \quad  z_2 > 1    
\end{array}
\right. 
\nonumber \\
p_3 & = & \left\{
\begin{array}{@{\,}ll} 
z_3^2 + (1-s_m)^2 \sin^2 \theta  &  {\rm for} \quad  z_3 < 0  \\ 
(1- s_m)^2 \sin^2 \theta  & {\rm for} \quad  0 \le z_3 \le 1  \\ 
 (1-z_3)^2 + (1-s_m)^2 \sin^2 \theta  & {\rm for} \quad  z_3 > 1    
\end{array}
\right. 
\end{eqnarray} 
Then, we define $d_{min}^2$ by the minimum of $p_j$ for 
$j=0, 1, 2, 3$.  

\par 
Finally, we discuss the case  when $\theta=0$ or $\pi$. 
When $\theta=0$, we define $D_{min}^2$  by 
\begin{equation}
D_{min}^2 = \left \{  
       \begin{array}{@{\,}lll}
           1+ 2 \alpha+ \gamma & {\rm for} \quad  \mbox{$\alpha < -1$} \\
           \gamma-\alpha^2  & {\rm for} \quad 
\mbox{$-1 \le$ $\alpha$ $\le$ 2} \\
           1-2\alpha+\gamma   & {\rm for} \quad \mbox{$\alpha > 2$}
       \end{array}
      \right. 
\end{equation}
When $\theta$=$\pi$, we define it by 
\begin{equation}
D_{min}^2 = \left \{  
       \begin{array}{@{\,}lll}
           \gamma & {\rm for} \quad  \mbox{$\alpha < 0$} \\
           \gamma-\alpha^2  & {\rm for} \quad \mbox{$0 \le \alpha \le 2$} \\
           4-4\alpha+\gamma   & {\rm for} \quad \mbox{$\alpha > 2$}
       \end{array}
      \right. 
\end{equation}
If $D_{min}$ is greater than $2r$, we have no overlap; 
otherwise there is an overlap. 

\newpage

{\bf Figure Captions}
\vskip 0.6cm 

\vspace{0.3cm}
\par \noindent 
Fig. 1:  Trivial knot,  prime knot $3_1$ (trefoil knot), and  composite knot $3_1 \# 3_1$.    
        Knot $3_1 \# 3_1$ is given by a product of two prime knots $3_1$. 

\vspace{1cm}
\par \noindent 
Fig. 2:  Polygonal knot equivalent to trefoil knot $3_1$. 

\vspace{0.8cm}
\par \noindent 
Fig. 3: Dimerization scheme for constructing self-avoiding walks of length 100;
 chains of lengths 12 and 13 are generated by 
a direct methods; then, chains of 25, 50 and 100  are 
given by the dimerization method, systematically. 

\vspace{0.8cm}
\par \noindent 
Fig. 4: Unknotting probability $P_0(N,r)$ versus number $N$ of polygonal nodes 
of the cylindrical self-avoiding polygons constructed 
by the dimerization method.
 Numerical estimates of $P_0(N,r)$ are shown 
for the following three values of $r$ 
by black circles, black triangles and black diamonds, respectively: 
  (a) $r$=0.001, 0.003 and 0.005; 
 (b) $r$=0.002, 0.004, and  0.006.  
Error bars denote their standard deviations.
Number  $N$ of nodes are given by 51, 151 and $100 j+1$ for $j=1, 2, \ldots,10$.  

\vspace{0.8cm}
\par \noindent 
Fig. 5: Knotting probability $P_K(N,r)$ of the cylindrical 
self-avoiding polygons constructed 
by the dimerization method
versus number  $N$ of polygonal nodes  for nontrivial prime knots. 
Numerical estimates of $P_K(N,r)$ for $K$=$3_1$, $4_1$ and $5_1$ 
are shown by black circles, black triangles and black diamonds, respectively, 
for the following values of $r$:  (a) $r$=0.001; (b) $r$=0.003; (c) $r$=0.005.  
Error bars denote their standard deviations.
Number $N$ of nodes are given by 51, 151 and $100j+1$ for $j=1,\cdots,10$.

\vspace{0.8cm} 
\par \noindent 
Fig. 6:  Knotting probability $P_K(N,r)$ of  the cylindrical self-avoiding polygons constructed 
by the dimerization method
versus number $N$ of polygonal nodes for  two knots with five crossings. 
Numerical estimates of $P_K(N,r)$  for $K$=$5_1$ and $5_2$
are shown by black circles and  black triangles, respectively, 
for the following values of $r$: (a) $r$=0.001; (b) $r$=0.003; (c) $r$=0.005.  
Error bars depicted in Fig. 6 are given simply by the sum of 
the statistical fluctuation of the number $M_K$ 
and that of the statistical weight of the ring-dimerization procedure.
Number  $N$ of nodes are given by 51, 151, and  $100j+1$ for $j=1, 2, \cdots,10$.  

\vspace{0.8cm}
\par \noindent 
Fig. 7: Knotting probability $P_K(N,r)$ of  the cylindrical self-avoiding polygons constructed 
by the dimerization method
versus number $N$ of polygonal nodes for two composite  knots.  
Numerical estimates of $P_K(N,r)$ for 
$K$ = $3_1\sharp3_1$ and $3_1\sharp3_1\sharp3_1$  
are shown by black circles and  black triangles, respectively, 
for the following values of $r$: 
(a) $r$=0.001; (b) $r$=0.003;  (c) $r$=0.005.  
Error bars denote their standard deviations.
Number $N$ of nodes are given by 51, 151 and $100j+1$ for $j=1, 2, \cdots,10$ .  

\vspace{0.8cm} 
\par \noindent 
Fig. 8: Semi-logarithmic plot of characteristic length 
$N_K(r)$ versus cylinder radius $r$
of  the cylindrical self-avoiding polygons constructed 
by the dimerization method.  
Numerical estimates of $N_K(r)$  for $K$=0 and $3_1$
listed in Table 2 are depicted 
by black circles and  black triangles, respectively, 
 together with their error bars, with the values of 
$r$ from 0.0 to 0.007 by 0.001. 
All  the black triangles (knot $3_1$) 
are slightly shifted rightward by 0.0001 
for graphical convenience.   
The straight fitting line is determined by the least square method. 

\vspace{0.8cm}
\par \noindent
Fig. 9: Semi-logarithmic plot of characteristic length $N_K(r)$ 
versus cylinder radius $r$ of  
the cylindrical self-avoiding polygons constructed 
by the dimerization method.
Numerical estimates of $N_K(r)$  for 
$K$=0, $3_1$, $4_1$, and $5_1$ 
listed in Table 2 are depicted 
by black circles and  black triangles, black diamonds and black crosses,
 respectively, for the values of $r$ from 0.001 to 0.007 by 0.001, 
 together with their error bars.
Black triangles ($3_1$), black diamonds ($4_1$) and 
black crosses ($5_1$) are shifted rightward by 
0.0001, 0.0002 and 0.0003, respectively.   

\vspace{0.8cm}
\par \noindent
Fig.10:  Probability $P_{knotted}(N,r)$ of being knotted 
versus cylindrical diameter 2$r$ for the hedge-hog and dimerization methods.
Numerical estimates of $P_{knotted}(N,r)$ 
of the hedge-hog method for $N$ = 20 and 30 
are shown by black circles and  black triangles, respectively.
Numerical estimates of $P_{knotted}(N,r)$ of 
the dimerization method for $N$ = 21 and 31 
are shown by black diamonds and black crosses, respectively. 
Error bars denote their standard deviations.

\vspace{0.8cm}
\par \noindent
Fig. 11: The ratio $P_K(N,r)$/$P_K(N,0)$ versus number $N$ of polygonal nodes  
of  the cylindrical self-avoiding polygons constructed 
by the dimerization method. 
Numerical estimates of $P_K(N,r)$/$P_K(N,0)$ for $r$ = 0.001, 0.005 and 0.007 
are shown by black circles, black triangles and black diamonds, respectively, 
for the following knots: (a) trivial knot; (b) trefoil knot $3_1$.

\twocolumn
\begin{table}
\caption{Values of the determinant of  knot $\mid \Delta_K(-1)\mid $  
and the second Vassiliev invariant $\it{v_2}$(K) for some simple knots.}
\label{inv}
\begin{tabular}{*{3}{c}}
  \hline 
   Knot K & $\mid \Delta_K(-1)\mid$  & $\it{v_2}$(K) \\
  \hline
    0    & 1 &  0  \\
   $3_1$ & 3 & -12 \\
   $4_1$ & 5 & 12  \\
   $5_1$ & 5 & -36 \\
   $5_2$ & 7 & -24 \\
   $3_1 \sharp 3_1$ & 9 & 24 \\
   $3_1 \sharp 3_1 \sharp 3_1$ & 27 & -36 \\ 
  \hline
\end{tabular}
\end{table}
\footnotesize

\begin{table}
\caption{Fitting parameters $m_K(r)$, $C_K(r)$, $N_K(r)$ to the cylindrical self-avoiding polygons constructed by the dimerization method}
\label{para}
\begin{center}
\begin{tabular}{*{8}{c}}
 \hline
 trivial & r=0.001 & r=0.002 & r=0.003 & r=0.004 & r=0.005 & r=0.006 & 
r=0.007\\
   \hline
  $C_K(r)$ & 1.055$\pm$0.065 & 1.048$\pm$0.006 & 1.066$\pm$0.061 & 1.034$\pm$0.060 & 1.001$\pm$0.058 & 1.034$\pm$0.057 & 1.024$\pm$0.055 \\

  $N_K(r)$ & 278$\pm$14 & 298$\pm$15 & 315$\pm$16 & 341$\pm$17 & 377$\pm$20 & 388$\pm$20 & 417$\pm$21 \\

 $m_K(r)$ & -0.004$\pm$0.034 & -0.002$\pm$0.03 & 0.0010$\pm$0.031 &-0.004$\pm$0.030 &  -0.019$\pm$0.028 & 0.001$\pm$0.027 & -0.006$\pm$0.025 \\

 $\chi^2$ & 3.0 & 2.7 & 1.2 & 2.1 & 2.0 & 2.2 & 1.7 \\
 \hline
\end{tabular}

\begin{tabular}{*{8}{c}}
 \hline
   $3_1$ & r=0.001 & r=0.002 & r=0.003 & r=0.004 & r=0.005 & r=0.006 & r=0.007\\
   \hline
   $C_K(r)$ & 0.625$\pm$0.021 &0.638$\pm$0.020 & 0.655$\pm$0.061 & 0.669$\pm$0.019 & 0.675$\pm$0.020 & 0.687$\pm$0.021 & 0.678$\pm$0.020 \\

   $N_K(r)$ & 264$\pm$21 & 280$\pm$22 & 297$\pm$24 & 314$\pm$26 & 341$\pm$30 & 375$\pm$36 & 390$\pm$38 \\

  $m_K(r)$ & 1.06$\pm$0.10 & 1.091$\pm$0.100 & 1.098$\pm$0.097 & 1.128$\pm$0.097 & 1.082$\pm$0.095 & 1.048$\pm$0.094 & 1.072$\pm$0.095 \\
 
  $\chi^2$ & 0.9 & 1.7 & 0.5 & 1.0 & 1.9 & 0.7 & 2.6 \\
   \hline
\end{tabular}

\begin{tabular}{*{8}{c}}
 \hline
   $4_1$ & r=0.001 & r=0.002 & r=0.003 & r=0.004 & r=0.005 & r=0.006 & r=0.007\\  
  \hline 
   $C_K(r)$ & 0.124$\pm$0.009 & 0.125$\pm$0.009 & 0.125$\pm$0.009 & 0.128$\pm$0.012 & 0.125$\pm$0.009 & 0.126$\pm$0.009 & 0.131$\pm$0.008 \\

   $N_K(r)$ & 236$\pm$40 & 259$\pm$47 & 286$\pm$56 & 348$\pm$75 & 348$\pm$79 & 347$\pm$79 & 334$\pm$70 \\

   $m_K(r)$ & 1.269$\pm$0.236 & 1.240$\pm$0.244 & 1.176$\pm$0.24 &0.978$\pm$0.218 &  1.118$\pm$0.239 & 1.148$\pm$0.243 & 1.247$\pm$0.236 \\
   
   $\chi^2$ & 2.6 & 2.3 & 1.0 & 2.0 & 1.7 & 2.3 & 1.4 \\
   \hline
\end{tabular}

\begin{tabular}{*{8}{c}}
 \hline
   $5_1$ & r=0.001 & r=0.002 & r=0.003 & r=0.004 & r=0.005 & r=0.006 & r=0.007\\  
  \hline 
   $C_K(r)$ & 0.035$\pm$0.006 & 0.041$\pm$0.005 & 0.040$\pm$0.005 & 0.037$\pm$0.005 & 0.034$\pm$0.007 & 0.036$\pm$0.004 & 0.035$\pm$0.005 \\

   $N_K(r)$ & 198$\pm$56 & 246$\pm$72 & 311$\pm$116 & 265$\pm$87 & 222$\pm$62 & 326$\pm$129 & 425$\pm$222 \\

 $m_K(r)$ & 1.612$\pm$0.486 & 1.335$\pm$0.395 & 1.103$\pm$0.428 & 1.380$\pm$0.436 &1.797$\pm$0.452 &  1.306$\pm$0.448 & 1.118$\pm$0.473  \\
  
   $\chi^2$ & 1.6 & 3.6 & 0.7 & 2.1 & 2.6 & 0.9 & 0.8 \\
  \hline
\end{tabular}

\begin{tabular}{*{8}{c}}
 \hline
   $5_2$ & r=0.001 & r=0.002 & r=0.003 & r=0.004 & r=0.005 & r=0.006 & r=0.007\\  
  \hline 
   $C_K(r)$ & 0.068$\pm$0.007 & 0.072$\pm$0.007 & 0.068$\pm$0.006 & 0.064$\pm$0.006 & 0.060$\pm$0.006 & 0.062$\pm$0.006 & 0.061$\pm$0.006 \\

   $N_K(r)$ & 226$\pm$49 & 230$\pm$51 & 283$\pm$73 & 278$\pm$73 & 322$\pm$96 & 330$\pm$101 & 290$\pm$81 \\

 $m_K(r)$ & 1.401$\pm$0.331 & 1.404$\pm$0.340 & 1.234$\pm$0.324 & 1.377$\pm$0.347 &1.317$\pm$0.340 &  1.296$\pm$0.348 & 1.418$\pm$0.367  \\
  
   $\chi^2$ & 1.3 & 2.0 & 1.2 & 1.6 & 2.0 & 1.3 & 1.0 \\
  \hline
\end{tabular}
\end{center}
\end{table}

\newpage
\begin{table}
\begin{center}
\begin{tabular}{*{8}{c}}
 \hline
   $3_1\sharp3_1$ & r=0.001 & r=0.002 & r=0.003 & r=0.004 & r=0.005 & r=0.006 & r=0.007\\  
  \hline
  $C_K(r)$ & 0.166$\pm$0.026 & 0.171$\pm$0.029 & 0.161$\pm$0.032 & 0.187$\pm$0.032 & 0.184$\pm$0.034 & 0.173$\pm$0.038 & 0.161$\pm$0.040 \\

 $N_K(r)$ & 259$\pm$36 & 273$\pm$41 & 278$\pm$43 & 326$\pm$57 & 318$\pm$52 & 328$\pm$58 & 325$\pm$59 \\
  
 $m_K(r)$ & 2.150$\pm$0.238 & 2.178$\pm$0.248 & 2.311$\pm$0.260 & 2.102$\pm$0.251 & 2.225$\pm$0.245 & 2.327$\pm$0.266 & 2.448$\pm$0.282 \\

 $\chi^2$ & 1.9 & 0.9 & 1.4 & 1.9 & 2.0 & 1.1 & 1.6 \\
  \hline
\end{tabular}

\begin{tabular}{*{8}{c}}
 \hline
   $3_1\sharp3_1\sharp3_1$ & r=0.001 & r=0.002 & r=0.003 & r=0.004 & r=0.005 & r=0.006 & r=0.007\\  
  \hline
  $C_K(r)$ & 0.025$\pm$0.020 & 0.023$\pm$0.021 
& 0.049$\pm$0.038 & 0.021$\pm$0.022 & 0.044$\pm$0.039 & 0.030$\pm$0.036 & 0.026$\pm$0.029 \\  
   $N_K(r)$ & 249$\pm$77 & 244$\pm$78 & 356$\pm$152 
& 260$\pm$90 & 362$\pm$163 & 330$\pm$161 & 318$\pm$138 \\
   $m_K(r)$ & 3.316$\pm$0.689 & 3.443$\pm$0.741 & 2.838$\pm$0.672 & 3.602$\pm$0.783 & 2.976$\pm$0.719 & 3.358$\pm$0.888 & 3.478$\pm$0.804 \\   
   $\chi^2$ & 0.2 & 1.0 & 0.5 & 0.7 & 0.8 & 1.3 & 0.9 \\
   \hline
\end{tabular}
\end{center}
\end{table}


\begin{thebibliography}{[99]}

\bibitem{Dean} F.B. Dean, A. Stasiak, T. Koller 
and N.R. Cozzarelli: J. Biol. Chem. {\bf 260} (1985) 4795; \\ 
S.A. Wasserman, J.M. Duncan and N.R. Cozzarelli:  
Science {\bf 229} (1985)171; Science {\bf 232} (1986) 951. 

\bibitem{Shishido} K. Shishido, N. Komiyama and S. Ikawa: 
J. Mol. Biol. $\bf{195}$ (1987) 215.

\bibitem{Walba} D.M. Walba, Tetrahedron: $\bf{41}$ (1985) 3161.

\bibitem{Delbruck} M. Delb$\ddot{\rm u}$ck:  
{\it Mathmatical Problems in the Biological Sciences}, 
edited by R.E. Bellman: Proc. Symp. Appl. Math $\bf{14}$ (1962) 55.

\bibitem{Frisch-Wasserman}
 H.L. Frisch and E. Wasserman:   
J. Amer. Chem. Soc. {\bf 83} (1961) 3789.


\bibitem{Vologodskii1} A.V. Vologodskii, A.V. Lukashin, M.D. Frank-Kamenetskii, and V.V. Anshelevich:   
Sov. Phys. JETP {\bf 39} (1974) 1059. 

\bibitem{desCloizeaux} J. des Cloizeaux and M.L. Mehta: 
J. Phys. (Paris) $\bf{40}$ (1979) 665.

\bibitem{Michels} J.P.J. Michels and F.W. Wiegel: 
Phys. Lett. A $\bf{90}$ (1982) 381.

\bibitem{LeBret} M. Le Bret: Biopolymers $\bf{19}$ (1980) 619.

\bibitem{Chen} Y.D. Chen: J. Chem. Phys. $\bf{74}$ (1981) 2034; J. Chem. Phys. $\bf{75}$, 2447 (1981); J. Chem. Phys. $\bf{75}$ (1981) 5160.

\bibitem{Klenin} K.V. Klenin, A.V. Vologodskii, V.V. Anshelevich, A.M. Dykhne and M.D. Frank-Kamenetskii: J. Biomol. Struct. Dyn. $\bf{5}$(1988) 1173.


\bibitem{Janse} E.J. Janse van Rensburg and S.G. Whittington: J.Phys. A $\bf{23}$ (1990) 3573. 

\bibitem{Koniaris} K. Koniaris and M. Muthukumar: 
Phys. Rev. Lett. $\bf{66}$ (1991) 2211.


\bibitem{DeguchiJKTR} T. Deguchi and K. Tsurusaki: 
J. Knot Theory and Its Ramifications $\bf{3}$ (1994) 321. 


\bibitem{Deguchi95} T. Deguchi and K. Tsurusaki: 
 {\it Geometry and Physics}, Lect. Notes in Pure and
Applied Math. Series/184, ed. by
J.E. Andersen, J. Dupont, H. Pedersen, and A. Swann, 
(Marcel Dekker Inc., Basel Switzerland, 1997),  pp. 557-565. 
(the Proceedings of {\it Geometry and Physics}, Institute of Mathematics, 
University of Aarhus, 18th-27th July, 1995, Aarhus, Denmark.) 

\bibitem{DeguchiRevE} T. Deguchi and K. Tsurusaki: 
Phys. Rev. E $\bf{55}$ (1997) 6245.

\bibitem{Orlandini} E. Orlandini, M.C. Tesi, E.J. Janse van Rensburg and S.G. Whittington:
J. Phys. A: Math. Gen. {\bf 31} (1998) 5953. 


\bibitem{Grosberg-book} A.Yu. Grosberg and A.R. Khokhlov: 
{\it Statistical Physics of Macromolecules}, AIP Press, 1994.    


\bibitem{Rybenkov} V.V. Rybenkov, N.R. Cozzarelli and A.V. Vologodskii:  
Proc. Natl. Acad. Sci. USA {\bf 90} (1993) 5307.  

\bibitem{Shaw}  
S.Y. Shaw and J.C. Wang: 
Science {\bf 260} (1993) 533.  


\bibitem{Stiger} D. Stiger: Biopolymer $\bf{16}$ (1977) 1435.

\bibitem{Brian} A.A. Brian, H.L. Frisch and  L.S. Lerman: Biopolymer $\bf{20}$, (1981) 1305.

\bibitem{OSF} 
T. Odijk: J. Poly. Sci.: Polym. Phys. {\bf 15} (1977) 477; 
  J. Skolnick and M. Fixman: Macromolecules {\bf 10} (1977) 944.  


\bibitem{PLA} M.K. Shimamura and T. Deguchi: 
preprint cond-mat/0008268, Phys. Lett. A $\bf{274}$ (2000) 184.

\bibitem{DeguchiPLA} 
T. Deguchi and K. Tsurusaki: Phys. Lett. A $\bf{174}$ (1993) 29; \\
See also, M. Wadati, T. Deguchi and Y. Akutsu: Phys. Reports {\bf 180}
(1989) 247; V.G. Turaev: Math. USSR Izvestiya {\bf 35} (1990) 411.   

\bibitem{Polyak} M. Polyak and O. Viro: Int. Math. Res. Not. No.11 (1994) 445.



\bibitem{Madras} N. Madras and G. Slade: {\it The Self-Avoiding Walk}, 
(Birkh{\"a}user, Boston, 1993), \S 9.3.2. 


\bibitem{deGennes} P.G. de Gennes: {\it Scaling Concepts of Polymer Physics} 
(Cornel University Press, Ithaca, 1979).  

\bibitem{Nechaev} A. Grosberg and S. Nechaev: 
J. Phys. A: Math. Gen. $\bf{25}$ (1992) 4659.

\end{thebibliography}
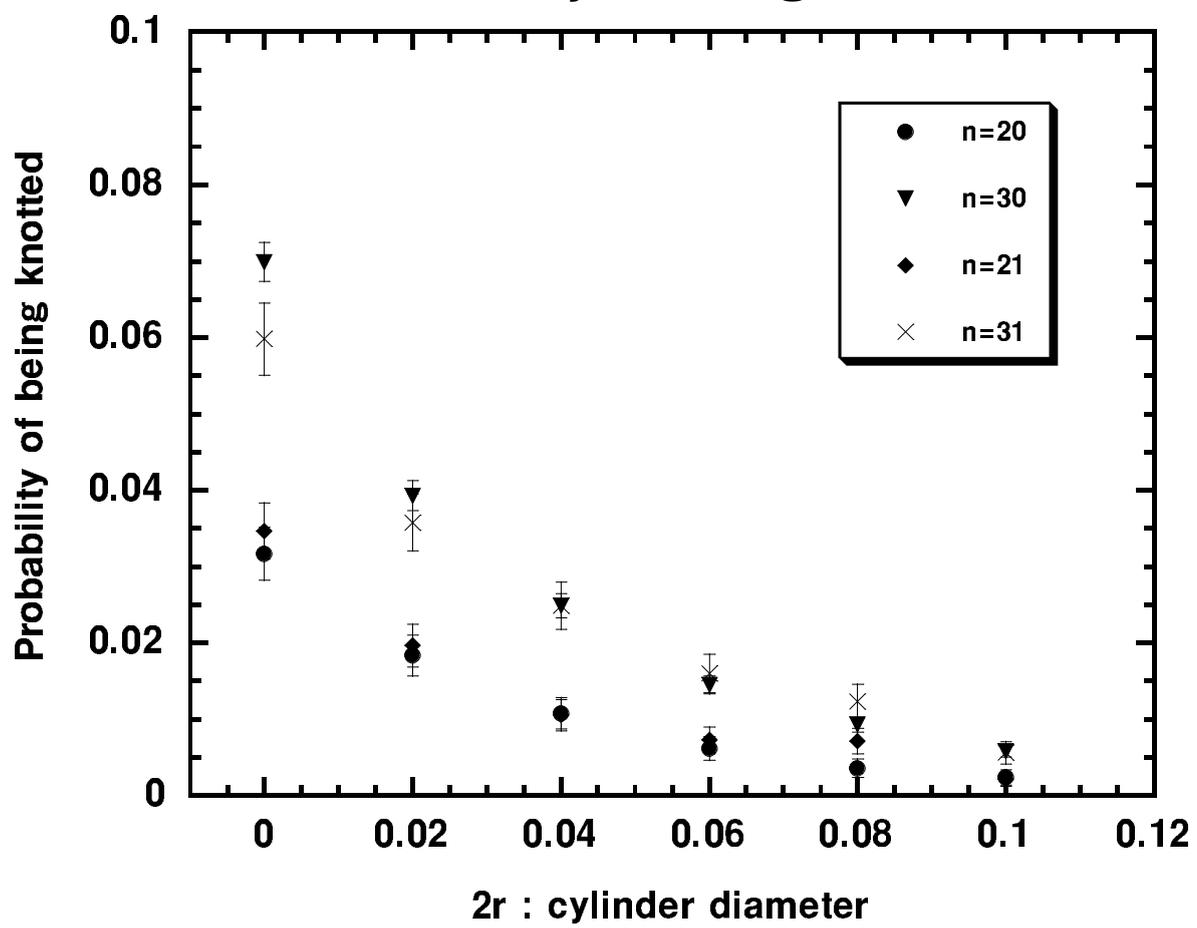
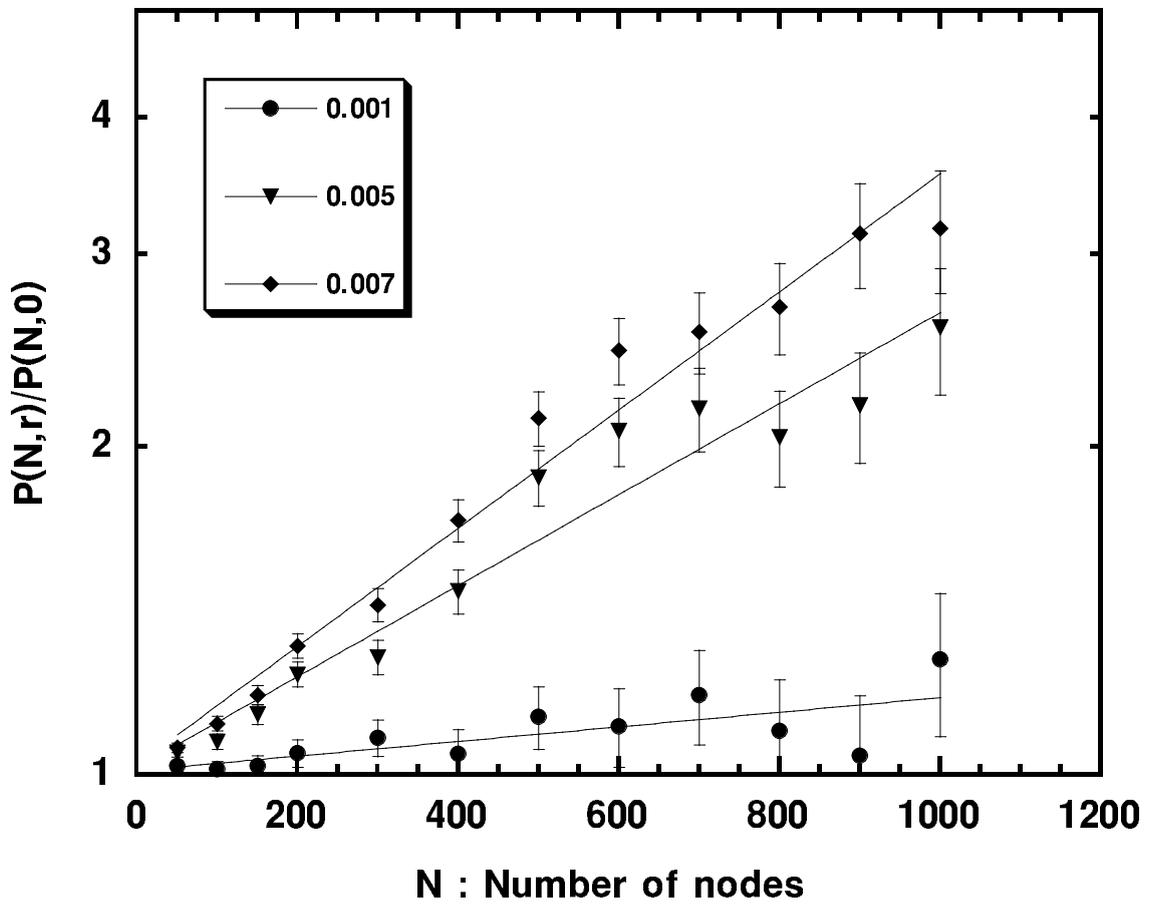
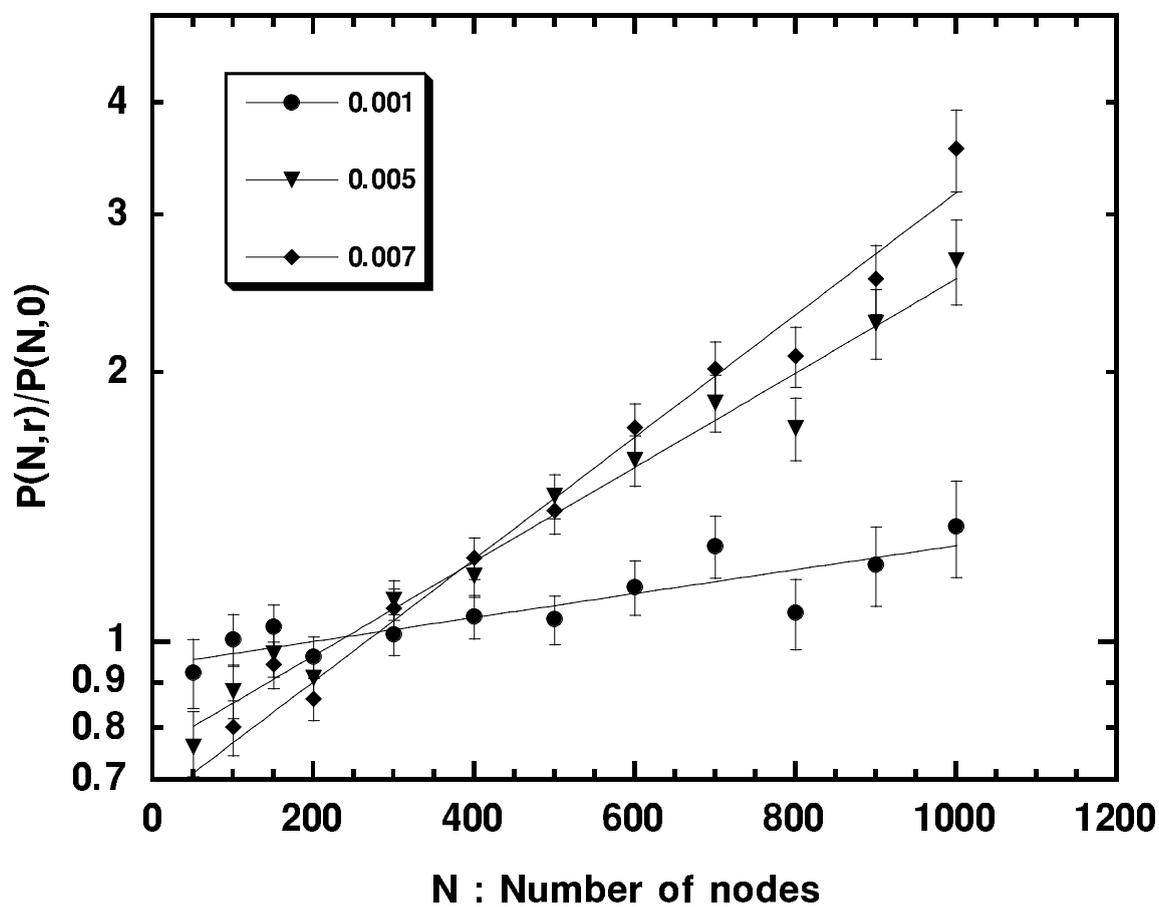
\end{document}